\documentstyle[preprint,pra,aps]{revtex}
\tighten

\begin{document}

\preprint{Preprint QMW--PH--95--35,
          submitted to Phys.~Rev.~A.}

\title{Quantum state diffusion with a moving basis: computing \\
          quantum-optical spectra}

\author{R\"udiger Schack, Todd A. Brun, and Ian C. Percival}
\address{Department of Physics, Queen Mary and Westfield College,\\
University of London, Mile End Road, London E1 4NS, UK}
\maketitle

\begin{abstract}
Quantum state diffusion (QSD) as a tool to solve quantum-optical master
equations by stochastic simulation can be made several orders of magnitude more
efficient if states in Hilbert space are represented in a {\it moving basis\/}
of excited coherent states. The large savings in computer memory and time are
due to the localization property of the QSD equation. We show how the method
can be used to compute spectra and give an application to second
harmonic generation.
\end{abstract}

\vskip2cm

Stochastic simulation of trajectories of Hilbert-space vectors or {\it quantum
trajectories\/} has proved to be a powerful new method for the solution of
master equations in quantum optics. The main advantage of stochastic simulation
methods over direct numerical solution of the master equation stems from the
fact that far less computer memory is needed to store a Hilbert-space vector
than to store a density operator. This advantage often far outweighs the main
disadvantage of stochastic simulation, namely that many trajectories have to be
added to obtain good statistics.

There are two main approaches to quantum trajectories in quantum optics: the
relative state method {\cite{Carmichael1993b,Dalibard1992,Gardiner1992}} and
quantum state diffusion (QSD) {\cite{Gisin1992c}}.  In the relative
state method, quantum trajectories are conditional on measurement outcomes.
Consequently, a single master equation can be ``unraveled'' into many different
stochastic equations, each corresponding to a different measurement scheme. For
a given problem, one is free to choose the unraveling that results in the
fastest algorithm. This flexibility is a major strength of the relative state
method, when the measurement process is the main interaction with the
environment. The relative state method is limited, however, to problems with
few degrees of freedom and to relatively small photon numbers. For two or more
field modes and large photon numbers, the number of basis states needed for a
numerical representation of a Hilbert-space vector can become very large,
leading to prohibitive computing times.

For the QSD method, on the other hand, there is a unique correspondence between
master equations and stochastic equations
{\cite{Gisin1992c,Diosi1988a}}. Although it has been shown that, under special
circumstances, the QSD equation is identical to the relative state equation
conditioned on heterodyne detection of photons {\cite{Wiseman1993a}}, in
general the QSD equation is not derived with respect to any specific
measurement scheme. The QSD method can be applied with equal ease whether or
not the main interaction with the environment is a measurement process.  For an
interpretation of the QSD equation, see {\cite{Gisin1992c}}.

QSD derives its strength as a computational tool from the {\it localization\/}
property of the QSD equation {\cite{Gisin1992c,Diosi1988c,%
Gisin1993b,Percival1994b,Percival1995a,Halliwell1995a}}.  We have shown
{\cite{Schack1995c}} that representing QSD trajectories in a moving basis of
excited coherent states instead of representing them in the usual number-state
basis can reduce the number of basis states needed---and thus the computing
time---by several orders of magnitude. We call the numerical method
based on localization {\it quantum state diffusion with a moving basis\/}
(MQSD). MQSD is particularly advantageous for problems with several degrees of
freedom and large photon numbers and can be used in regimes extending from the
highly quantum mechanical (few photons) to the semiclassical (very many
photons).

In this paper, we show how MQSD can be used to compute quantum-optical
spectra. Whereas computing spectra using the relative state method is a
well established procedure {\cite{Dum1992b}}, quantum state diffusion
has not been applied to the practical computation of spectra before. Here we
show that using MQSD for the computation of spectra leads to the same savings
as using MQSD for the computation of photon numbers as in {\cite{Schack1995c}},
and thus that it can be applied where the relative state method would be
impractical.

Many problems in quantum optics can be formulated in terms of a master equation
of Lindblad form \cite{Lindblad1976}
\begin{equation}
{d\over{dt}}\,\hat\rho = {\cal L}\hat\rho \equiv
 -{i\over\hbar}[\hat H,\hat\rho] +
 \sum_m\left(\hat L_m\hat\rho \hat L_m^\dagger
 - \case1/2 \hat L_m^\dagger\hat L_m\hat\rho   - \case1/2\hat\rho
 \hat L_m^\dagger\hat L_m\right) \;,
\label{eqmaster}
\end{equation}
where $\hat\rho$ is the system density operator, $\hat H$ is the system
Hamiltonian, and the $\hat L_m$ are
Lindblad operators which are generally not Hermitian and
represent the effect of the environment on the system
in the Markov approximation.  The QSD equation \cite{Gisin1992c} derived from
Eq.~(\ref{eqmaster}) is a nonlinear stochastic differential equation for a
normalized state vector $|\psi\rangle$:
\begin{eqnarray}
|d\psi\rangle &=& -{i\over\hbar} \hat H \,|\psi\rangle dt
  + \sum_m\left(\langle \hat L_m^\dagger\rangle \hat L_m
  - \case1/2 \hat L_m^\dagger \hat L_m
  - \case1/2\langle \hat L_m^\dagger\rangle
  \langle\hat L_m\rangle\right) |\psi\rangle dt \nonumber \\
&&+\sum_m \left(\hat L_m -\langle \hat L_m\rangle\right)
          \,|\psi\rangle d\xi_m \;.
\label{eqqsd}
\end{eqnarray}
The first sum in (\ref{eqqsd}) represents the deterministic drift of the state
vector due to the environment, and the second sum the random
fluctuations. Angular brackets denote the quantum expectation $\langle\hat
G\rangle = \langle \psi|\hat G|\psi\rangle$ of the operator $\hat G$ in the
state $|\psi\rangle$. The $d\xi_m$ are independent complex differential
Gaussian random variables satisfying the conditions
\begin{equation}
{\rm M} d\xi_m = {\rm M} d\xi_n d\xi_m = 0\;,\;\;\;
{\rm M} d\xi_n^* d\xi_m = \delta_{nm}dt \;,
\end{equation}
where {\rm M} denotes the ensemble mean. The density operator is given by the
mean over the projectors onto the quantum states of the ensemble:
\begin{equation}
\hat\rho = {\rm M}|\psi\rangle\langle\psi| \;.
\end{equation}
If the pure states of the ensemble satisfy the QSD equation (\ref{eqqsd}), then
the density operator satisfies the master equation (\ref{eqmaster}).

A crucial property of the QSD equation is localization.  The Schr\"odinger
evolution of an isolated system usually produces {\it de}localization or
dispersion.  According to the Schr\"odinger equation, initially localized wave
packets become highly delocalized except under very special circumstances.
This effect is also present in the first, Hamiltonian term in the QSD
equation~(\ref{eqqsd}).  The environment terms in the QSD equation, however,
have the opposite effect; they tend to make the wave packet narrower.  The
theory of localization is treated in {\cite{Gisin1992c,Percival1994b}}.
Numerical simulations
{\cite{Gisin1992c,Gisin1993b,Schack1995c,Garraway1994a}} show the competition
between the delocalizing effect of the Hamiltonian and the localizing effect of
the environment. The net effect is often a very well localized wave packet.

One important consequence of the localization of quantum trajectories is that,
by continually changing the basis, it is often possible to reduce the number of
basis states needed to represent the wave packet by several orders of
magnitude.  If a wave packet is localized about a point $(q,p)$ in phase space
far from the origin, it requires many number states $|n\rangle$ to represent
it.  But relatively few {\it excited coherent states\/} $|q,p,n\rangle = \hat
D(q,p)|n\rangle$, are needed, with corresponding savings in computer storage
space and computation time.  The operator $\hat D(q,p)$ is the usual coherent
state displacement operator {\cite{Louisell1973}},
\begin{equation}
\hat D(q,p) = \exp {i\over\hbar} \biggl( p\hat Q - q\hat P \biggr) \;,
\end{equation}
where $\hat Q$ and $\hat P$ are the position and momentum operators.
The separation of the representation into a classical part $(q,p)$ and
a quantum part $|q,p,n\rangle$ is called the {\it moving basis\/}
{\cite{Schack1995c}} or, as in
\cite{Percival1995a}, the {\it mixed\/} representation.

The QSD equation~(\ref{eqqsd}) can contain both localizing and delocalizing
terms.  Nonlinear terms in the Hamiltonian tend to spread the wave function in
phase space, whereas the Lindblad terms localize. Accordingly, the width of the
wave packets varies along a typical trajectory. We use this to reduce the
computing time even further by dynamically adjusting the number of basis
vectors. Details about the implementation of MQSD can be found
in {\cite{Schack1995c}}.

We now turn to the computation of optical spectra. The method we use simulates
the measurement of a spectral component at offset frequency $\omega$ and is
similar in spirit to {\cite{Tian1992,Plenio1994,Hegerfeldt1995}}.
It involves an auxiliary field mode or {\it
filter\/} mode weakly coupled to a system operator $\hat c$, detuned by the
frequency $\omega$, and weakly damped with damping constant $\kappa$.  The
master equation describing the system coupled to the filter mode can be written
as
\begin{equation}
{d\over dt}\hat\rho = {\cal L}\hat\rho
  - {i\over\hbar}[\hat H_{\rm f},\hat\rho]
  + 2\kappa ( \hat b\hat\rho \hat b^\dagger
  - \case1/2\hat b^\dagger\hat b\hat\rho
  - \case1/2\hat\rho\hat b^\dagger\hat b ) \;,  \label{eqmastot}
\end{equation}
where the system superoperator ${\cal L}$ is defined as in~(\ref{eqmaster}),
$\hat b$ is the annihilation operator for the filter mode, and
\begin{equation}
\hat H_{\rm f} = \hbar\omega\,\hat b^\dagger \hat b
  + i\hbar\epsilon\,(\hat c \hat b^\dagger - \hat c^\dagger \hat b)
\end{equation}
is the interaction picture Hamiltonian describing the filter mode and its
coupling to the system with coupling constant $\epsilon$. As in a real
experiment, the spectral resolution is limited by the filter damping constant
$\kappa$.

Assuming that the coupling $\epsilon$ is so small that there is a negligible
probability for the excitation of more than one photon in the filter mode, we
can write the total density operator as
\begin{equation}
\hat\rho \simeq  \hat\rho_{00} \otimes |0\rangle\langle0|
          + \hat\rho_{01} \otimes |0\rangle\langle1|
          + \hat\rho_{10} \otimes |1\rangle\langle0|
          + \hat\rho_{11} \otimes |1\rangle\langle1| \;,
\end{equation}
where the $\hat\rho_{ij}$ operate in the system Hilbert space, and the
$|i\rangle\langle j|$ operate in the filter Hilbert space. By expanding the
solution to the master equation in powers of the coupling $\epsilon$, we obtain
the following approximate equations of motion for the system operators
$\hat\rho_{00}$ and $\hat\rho_{01}$ {\cite{BrunGisin}}---notice that these
equations do not
preserve the trace.
\begin{eqnarray}
{d\over dt}\hat\rho_{00} &=& {\cal L}\hat\rho_{00} + O(\epsilon^2) \;, \\
{d\over dt}\hat\rho_{01} &=& {\cal L}\hat\rho_{01}
   + i\omega\,\hat\rho_{01} + \epsilon\,\hat\rho_{00}\hat c^\dagger
   - \kappa\hat\rho_{01} + O(\epsilon^3) \;.
\end{eqnarray}
These equations can be solved formally to yield
\begin{eqnarray}
\hat\rho_{00}(t) &=& e^{{\cal L}t} \, \hat\rho_{00}(0) + O(\epsilon^2) \;, \\
\hat\rho_{01}(t) &=& \epsilon\,\int_0^t e^{{\cal L}(t-t')}
  \left( \hat\rho_{00}(t') \hat c^\dagger \right)
  e^{(i\omega-\kappa)(t-t')} \,dt' + O(\epsilon^3) \;. \label{eqrho01}
\end{eqnarray}

We can now obtain an approximation to the Fourier transform of the
stationary-state time correlation function $\langle\hat c^\dagger(0)\hat
d(\tau)\rangle_{\rm ss}$ for an arbitrary system operator $\hat d$ by computing
the expectation $\langle\hat b^\dagger\hat d\hat b\hat b^\dagger\rangle$ in the
stationary regime (by $\langle\cdots\rangle_{\rm ss}$ we denote the expectation
value in the stationary state):
\begin{eqnarray}
{\rm tr}\left(\hat\rho(t)\hat b^\dagger\hat d\hat b\hat b^\dagger\right)
&=& {\rm tr}\left[\hat b\hat b^\dagger\left(\hat\rho_{01}(t)\otimes|0\rangle
       \langle1|\right)\,\hat b^\dagger\hat d\,\right]
={\rm tr}\left[\left(\hat\rho_{01}(t)\otimes|0\rangle
       \langle0|\right)\hat d\,\right]
= {\rm tr}\left(\hat d\hat\rho_{01}(t)\right)               \nonumber\\
&\simeq&\epsilon\,\int_0^t {\rm tr}\left[\hat d\,e^{{\cal L}(t-t')}
  \left( \hat\rho_{00}(t') \hat c^\dagger \right) \right]
  e^{(i\omega-\kappa)(t-t')} \,dt'   \\
&=&\epsilon\,\int_0^t   \langle\hat c^\dagger(t')\hat d(t)\rangle
  e^{(i\omega-\kappa)(t-t')} \,dt'  \;, \nonumber
\end{eqnarray}
where the quantum regression theorem {\cite{Carmichael1993b,Lax1967}} has been
used in the last line.  In the limit $t\rightarrow\infty$, one obtains
\begin{equation}
\langle\hat b^\dagger\hat d\hat b\hat b^\dagger\rangle_{\rm ss}
\simeq \epsilon\,\int_0^\infty
 \langle\hat c^\dagger(0)\hat d(\tau)\rangle_{\rm ss}
 e^{(i\omega-\kappa)\tau} \,d\tau \;.
\label{eqbdbb}
\end{equation}
Similarly, one finds
\begin{equation}
\langle\hat b\hat b^\dagger\hat d\hat b\rangle_{\rm ss}
\simeq \epsilon\,\int_0^\infty
 \langle\hat d(\tau)\hat c(0)\rangle_{\rm ss}
 e^{(-i\omega-\kappa)\tau} \,d\tau \;.
\label{eqbbdb}
\end{equation}
By computing the expectation values $\langle\hat b^\dagger\hat d\hat b\hat
b^\dagger\rangle_{\rm ss}$ and $\langle\hat b\hat b^\dagger\hat d\hat
b\rangle_{\rm ss}$ for various choices of the system operators $\hat c$ and
$\hat d$ using straightforward simulation of the QSD equation, one can thus
obtain approximations for various spectral densities at the offset frequency
$\omega$. For each value of $\omega$, a new simulation is needed.

In the following, we apply the method outlined above to the problem of second
harmonic generation or frequency doubling. The system consists of two optical
modes of frequency $\omega_1$ and $\omega_2=2\omega_1$ which interact in a
cavity driven by a coherent external field with frequency $\omega_1$ and
amplitude $E$.  The Hamiltonian in the interaction picture is
\begin{equation}
\hat H = i\hbar E (\hat a_1^\dagger-\hat a_1)
       + i\hbar{\chi\over2}
             (\hat a_1^{\dagger2} \hat a_2 - \hat a_1^2 \hat a_2^\dagger) \;,
\end{equation}
where $\hat a_1$ and $\hat a_2$ are the annihilation operators of the two
cavity modes, and $\chi$ describes the strength of the nonlinear interaction
between them.  Damping of the two cavity modes is described by the Lindblad
operators $\hat L_1=\sqrt{2\gamma_1}\hat a_1$ and $\hat
L_2=\sqrt{2\gamma_2}\hat a_2$. The factors of $\sqrt{2}$ are a consequence of
the convention used in the master equation (\ref{eqmaster}) which differs from
the usual convention in quantum optics.

The master equation for this problem first appeared in \cite{Drummond1981a}.
Earlier numerical treatments include
{\cite{Schack1995c,Doerfle1986,Savage1988,Schack1991c,Goetsch1993,Zheng1995}};
none of these compute spectra. The main difficulty in a numerical treatment of
second harmonic generation is due to the large size of the number-state basis
needed to represent a Hilbert-space vector. If $n_1$ number states are needed
for mode 1 and $n_2$ number states are needed for mode 2, then the total number
of basis states needed is $n_1n_2$, which can be very large. By using MQSD, and
thus exploiting the localization property of the QSD equation, this number
can be reduced significantly {\cite{Schack1995c}}. We are then in a position to
compute spectra with moderate numerical effort.

Of particular interest is the squeezing spectrum in the upconverted mode $\hat
a_2$ as observed in homodyne detection.  Squeezing in the upconverted mode in
second harmonic generation has been experimentally observed by Sizmann et
al.~{\cite{Sizmann1990a}}. The squeezing spectrum $S_{\theta,\kappa}(\omega)$
{\cite{Carmichael1993b}} is defined in terms of the quadrature operators
\begin{equation}
\hat A_\theta(t) = \frac{1}{2}
  \left[ \hat a_2(t)e^{-i\theta} + \hat a_2^\dagger(t)e^{i\theta} \right]
\end{equation}
and
\begin{equation}
\Delta\hat A_\theta(t) = \hat A_\theta(t) - \langle\hat A_\theta(t)\rangle
\end{equation}
as
\begin{equation}
S_{\theta,\kappa}(\omega)
 = 16\gamma_2 \int_0^\infty d\tau\,\cos(\omega\tau)e^{-\kappa\tau}
 \langle:\Delta\hat A_\theta(0)\Delta\hat A_\theta(\tau):\rangle_{\rm ss} \;.
\label{eqspectrum}
\end{equation}
The notation $\langle:\cdots:\rangle$ means normal ordering and time ordering
of the operator products, and a nonzero value of $\kappa$ results in a limited
spectral resolution. A few lines of algebra lead to
\begin{eqnarray}
\label{eqspeclong}
S_{\theta,\kappa}(\omega)
&=& 4\gamma_2\,{\rm Re}\biggl[ \int_0^\infty d\tau\,e^{-\kappa\tau}
 \left(e^{i\omega\tau}+e^{-i\omega\tau}\right)
 \left( e^{-2i\theta} \langle\hat a_2(\tau)\hat a_2(0)\rangle_{\rm ss}
 + \langle\hat a_2^\dagger(0)\hat a_2(\tau)\rangle_{\rm ss} \right) \\
&& - \frac{2\kappa}{\kappa^2+\omega^2}\,\left( e^{-2i\theta}
 \langle\hat a_2\rangle_{\rm ss}^2
 + \langle\hat a_2^\dagger\rangle_{\rm ss} \langle\hat a_2\rangle_{\rm ss}
 \right) \biggr] \nonumber \;.
\end{eqnarray}
All the quantities in this equation can be easily determined by simulation of
the QSD equation with the help of Eqs.~(\ref{eqbdbb})
and~(\ref{eqbbdb}).

In this paper, we do not attempt to reproduce the results of Sizmann et al.;
instead, we choose a set of parameter values leading to a limit cycle in the
corresponding classical dynamics. The numerical treatment of this regime has
proven to be particularly challenging {\cite{Doerfle1986}}.  Figure~\ref{fig}
shows the resulting spectrum Eq.~(\ref{eqspectrum}) with $\theta=\pi/2$. The
spectral density is positive throughout; there is no squeezing in this regime,
which agrees with the findings in {\cite{Doerfle1986}}. The stationary
expectation values required for the evaluation of Eq.~(\ref{eqspeclong}) were
generated by computing time averages in the stationary regime, assuming
ergodicity. The integration time was $T=100\kappa^{-1}$.
Since the filter decay time $\kappa^{-1}$ was chosen to be the
largest decay time in the system, the statistics resulting from an integration
time $T=n\kappa^{-1}$ corresponds roughly to the statistics resulting from
adding
$n$ trajectories.  The computing time was about 1 day per data point on a
standard Pentium PC running Linux.

In conclusion, we have shown that the method of quantum state diffusion with a
moving basis or MQSD can be used to compute quantum optical spectra, with
considerable savings in computing time and memory compared with earlier
methods.

We thank the EPSRC in the UK and the University of Geneva for essential
financial support, and G.~Alber, N.~Gisin, P.~Knight, M.~Plenio, W.~Strunz and
P.~Zoller for valuable communications.



\begin{figure}
\caption{Homodyne spectrum Eq.~({\protect\ref{eqspectrum}}) with $\theta=\pi/2$
for the upconverted mode in second harmonic generation. The frequency scale on
the plot is in units of the decay constant of the fundamental mode, $\gamma_1$.
The parameters are $E/\gamma_1=20$, $\chi/\gamma_1=0.4$, $\gamma_2/\gamma_1=1$,
$\kappa/\gamma_1=0.1$, $\epsilon/\gamma_1=10^{-6}$, corresponding to a limit
cycle regime of the classical dynamics.}
\label{fig}
\end{figure}

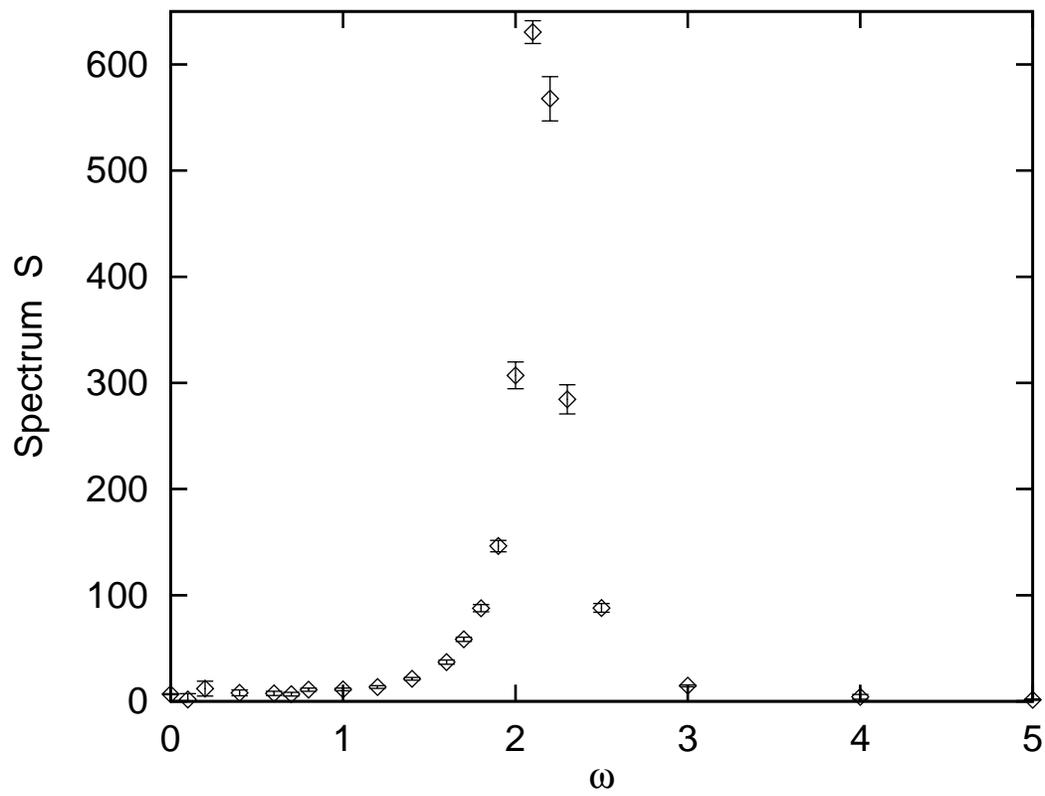
\end{document}